\documentclass[a4paper,11pt]{article}
\usepackage{amsmath}
\usepackage{amssymb}

\DeclareMathOperator{\re}{Re}

\DeclareMathOperator{\tr}{tr}
\DeclareMathOperator{\ch}{ch}

\begin{document}




\thispagestyle{empty}
\begin{titlepage}

\vspace*{-1cm}
\hfill \parbox{3.5cm}{BUTP-97/22 \\ hep-th/9708061} \\
\vspace*{1.0cm}

\begin{center}
  {\large {\bf \hspace*{-0.2cm}On the Spontaneous Identity
      \protect\vspace*{0.2cm} \\
      \protect\hspace*{0.2cm} of Chiral and Super Symmetry Breaking
      \protect\vspace*{0.2cm} \\
      \protect\hspace*{0.2cm} in Pure Super Yang Mills Theories}\footnote{Work
      supported in part by the Schweizerischer Nationalfonds.}  }
  \vspace*{3.0cm} \\

{\bf
    M. Leibundgut and P. Minkowski} \\
    Institute for Theoretical Physics \\
    University of Bern \\
    CH - 3012 Bern , Switzerland
   \vspace*{0.8cm} \\  

19 May 1998

\vspace*{4.0cm}

\begin{abstract}
\noindent
    We show that in supersymmetric pure Yang Mills theories with arbitrary
    simple gauge group, the spontaneous breaking of chiral fermionic and
    bosonic charge by the associated gaugino and gauge boson condensates
    implies the spontaneous breaking of supersymmetry by the condensate of the
    underlying Lagrangian density. The explicit breaking of the restricted
    fermionic charge through the chiral anomaly is deferred to a secondary
    stage in the elimination of infrared singularities or long range forces.
\end{abstract}
\end{center}

\end{titlepage}










\section{Introduction}
\label{sec:intro}

The question whether supersymmetry is spontaneously broken or not is of
fundamental importance. Many results concerning this problem have been derived
in the literature. We know that perturbative corrections do not break
supersymmetry.  What happens nonperturbatively is not yet clear since there is
no theory available to describe this regime.

A particular question is, whether or not there is a relation between the
breaking of chiral symmetry and the breaking of supersymmetry. In the case of
$N=1$ pure SYM theories this question has, besides other things, been
addressed in \cite{ven}.  The cited work concludes, that supersymmetry remains
unbroken after the breakdown of chiral symmetry.  This conclusion is
incomplete, since the effective action derived in this analysis does not have
a definite chiral weight. Potential $\theta$ angles do not relax, in which
case chiral symmetry breaking entails spontaneous CP violation.

We propose to use an appropriate thermodynamical limit of the pure SYM theory.
To explore the response of the theory to perturbations we introduce a
multiplet of external fields. This is analogous to measuring the hysteresis
lines of a ferromagnet exposed to an external magnetic field. In particular
one is forced to explicitly break susy by introducing an arbitrary complex
mass for the gaugino, corresponding to breaking the rotational invariance of
the ferromagnet by the external magnetic field. Note that every potential
spontaneous phenomenon necessarily entails a thermodynamical external field
extension.  Whether the spontaneous phenomenon actually occurs is a dynamical
question and is tantamount to follow the relaxation of the static external
symmetry breaking fields. The situation in QCD with $n$ massless quark flavors
\cite{PMeta} is identical in the sense that the intermediary external field
extension necessarily incorporates nonvanishing complex quark masses.

In the thermodynamic limit the $\theta$ angles relax and chiral symmetry
\cite{Crew} is restored. Therefore the different terms in the effective action
must have a definite weight under chiral rotations. This knowledge together
with the supersymmetric form of the effective action extended to include the
external sources suffices to derive consistency conditions for the formation
of gaugino and gauge boson condensates.  It then follows that in the limit of
vanishing gaugino mass, the gaugino condensate can not exist without a
corresponding condensate of the gauge bosons and vice versa. Since the gauge
boson condensate breaks susy, there is an intimate relation between chiral and
supersymmetry breaking. Assuming a confining mechanism similar to the one in
QCD, we conclude that susy must be broken. This result does not agree with the
conclusion obtained in \cite{witten_index}. However our derivation of the
thermodynamical limit involves the use of nontrivial boundary conditions,
whereas in \cite{witten_index} trivial boundary conditions are assumed.

This paper is organized in the following manner. In section \ref{sec:static}
we derive the thermodynamical limit of the pure SYM theory. The chiral anomaly
together with the relaxation of the $\theta$ angles is discussed in section
\ref{sec:chiral_anomaly}. Section \ref{sec:link_susybr_chiralbr} contains the
final link between chiral and susy breaking. We summarize our results in
section \ref{sec:conclusion} and give the derivation of the path measure in
the appendix.




\section{The thermodynamic limit of the $N=1$ SYM theory}
\label{sec:static}

In this section we discuss the thermodynamic limit of a pure super Yang Mills
theory with simple gauge group $G$.

In order to explore the hysteresis lines of the SYM theory in the thermodynamic
limit, we use a chiral multiplet of sources to drive the operators in the SYM
Lagrangian. It is convenient to write the action of the theory as bilinear in
the chiral external source superfield $J$ and the chiral superfield
\begin{equation}
  \label{eq:Def_Phi}
  \Phi = \frac{1}{4}\frac{1}{C_2(G)} \tr W^\alpha W_\alpha
\end{equation}
where $W_\alpha$ denotes the superspace field strength $W_\alpha=-\bar{D}^2
e^{-V}D_\alpha e^{V}$ of the vector superfield $V$. In the Wess-Zumino gauge
its components are
\begin{equation}
  \label{eq:super_fieldstrength_wz}
  W_\alpha(y,\vartheta)
  =\lambda_\alpha(y) + \vartheta_\alpha D(y)
  + \vartheta^\beta i F_{\alpha\beta}(y)
  + \vartheta^2 i D_{\alpha\dot{\alpha}}\bar{\lambda}^{\dot{\alpha}}(y)
\end{equation}
The component fields $\lambda^\alpha, F_{\alpha\beta} = \frac{1}{4}
(\sigma^{\left[\mu\right.}  \bar{\sigma}^{\left.\nu\right]})_{\alpha\beta}
F_{\mu\nu} $ and $D$ in eq.~(\ref{eq:super_fieldstrength_wz}) depend on $y^\mu
= x^\mu-\frac{i}{2}\vartheta\sigma^\mu\bar{\vartheta}$ and are in the adjoint
representation of the gauge group $G$,
\begin{equation}
  \label{eq:adj_rep}
  \lambda=\lambda^a T^a, \quad F_{\mu\nu}=F^a_{\mu\nu} T^a, \quad D=D^a T^a,
  \quad \tr T^a T^b = C_2(G) \delta^{ab}
\end{equation}
The action of the covariant derivative on the gaugino is
\begin{equation}
  \label{eq:gauge_cov_der}
  D_{\alpha\dot{\alpha}}\bar{\lambda}^{\dot{\alpha}}
  = \partial_{\alpha\dot{\alpha}}\bar{\lambda}^{\dot{\alpha}}
  + i \left[A_{\alpha\dot{\alpha}},\bar{\lambda}^{\dot{\alpha}}\right]
\end{equation}
where $A_{\alpha\dot{\alpha}} = \sigma^\mu_{\alpha\dot{\alpha}} A_\mu$ denotes
the gauge field. After rescaling the gaugino to its conventional normalization
$\lambda=\sqrt{2} \Lambda$, the chiral superfield $\Phi$ defined in
eq.~(\ref{eq:Def_Phi}) contains the $x$ dependent component fields
\begin{equation}
  \label{eq:phi}
  \begin{split}
    \Phi = 
    \vartheta^2 & \left(
      -\frac{1}{4}F^a_{\mu\nu}F^{a\mu\nu}
      +\frac{1}{2}\Lambda^a \overset{\leftrightarrow}{D}^{ab} \bar{\lambda}^b
      +\frac{i}{4} F^a_{\mu\nu}\tilde{F}^{a\mu\nu}
      + \frac{i}{4}\partial_\mu(\Lambda^a \sigma^\mu \bar{\Lambda}^a)
    \right) \\ 
    &+ \vartheta^\alpha \left(-\frac{1}{\sqrt{2}} \Lambda^{a \beta}
    F^a_{\beta\alpha} \right)
    + \frac{1}{2} \Lambda^{a \alpha}\Lambda^a_\alpha
  \end{split}
\end{equation}
In the sequel we will use the reparametrization
\begin{equation}
  \label{eq:phirep}
  \Phi = \vartheta^2 L(x) + \vartheta^\alpha \psi_\alpha(x) + \frac{1}{2} z(x)
\end{equation}
The source multiplet $J$ is a full chiral superfield,
\begin{equation}
  \label{eq:J}
  J = - \vartheta^2 m(x)
  + \vartheta^\alpha \eta_\alpha(x) + \frac{1}{2} j(x)
\end{equation}
All the variables $L, \psi$ and $z$ as well as their sources $j, \eta_\alpha$
and $m$ are a priori complex.

A generic bilinear in $\Phi$ and $J$ is
\begin{equation}
  \label{eq:general_bilinear}
  \Phi \cdot J =
  \vartheta^2 \left(\frac{1}{2}(jL-mz)-\psi^\alpha\eta_\alpha \right)
  + \frac{1}{2}\vartheta^\alpha(j\psi_\alpha+z\eta_\alpha)
  + \frac{1}{4} j z
\end{equation}
Since we are only interested in gaugino and gauge boson condensates, we set
the fermionic source $\eta_\alpha$ to zero in what follows. Projecting onto
the highest $\vartheta$-component leads to the generic external field
Lagrangian density
\begin{equation}
  \label{eq:L_J_generic}
  \mathcal{L}_J = \int d^2\vartheta (\Phi \cdot J + \text{h.c.})
  = \re(jL-mz)
\end{equation}
Choosing
\begin{equation}
  \label{eq:j_m_couplings}
  j(x)=\frac{1}{g^2(x)}-i\frac{\theta(x)}{8\pi^2}, \qquad m(x)=m_1(x)-i m_2(x)
\end{equation}
and rescaling the component fields according to
\begin{equation}
  \label{eq:rescaled_components}
  A \to g A, \qquad \Lambda \to g \Lambda, \qquad D \to g D
\end{equation}
we find the component fields of the generic Lagrangian
\begin{eqnarray}
  \label{eq:L_g_theta_generic}
  \mathcal{L}_J &=& 
  - \frac{1}{4g^2(x)} F^a_{\mu\nu} F^{a\mu\nu}
  + \frac{\theta(x)}{32\pi^2} F^a_{\mu\nu} \tilde{F}^{a\mu\nu} \nonumber \\
  && + \frac{1}{2}
  \Lambda^a i \overset{\leftrightarrow}{D}^{ab} \bar{\Lambda}^b
  + \frac{\theta(x)}{16\pi^2} 
  \partial_\mu(\Lambda^a\sigma^\mu\bar{\Lambda}^a)  \\
  && - \frac{1}{2} m_1(x) (\Lambda^{a\alpha}\Lambda^a_\alpha 
  + \bar{\Lambda}^a_{\dot{\alpha}}\bar{\Lambda}^{a\dot{\alpha}})
  - \frac{i}{2} m_2(x) (\Lambda^{a\alpha}\Lambda^a_\alpha 
  - \bar{\Lambda}^a_{\dot{\alpha}}\bar{\Lambda}^{a\dot{\alpha}}) \nonumber
\end{eqnarray}
The SYM Lagrangian with restored coupling constants is represented in the
analogous way
\begin{equation}
  \label{eq:L_0}
  \mathcal{L}_0 = \int d^2\vartheta (\Phi \cdot J_0 + \text{h.c.}), \qquad
  J_0 = \frac{1}{2} \left(
    \frac{1}{g_0^2} - i \frac{\theta_0}{8\pi^2} \right) = \text{const.} 
\end{equation}
The generating functional for the Green functions in the presence of general
$x$ dependent external sources $J$ is
\begin{equation}
  \label{eq:generating_functional}
  \mathcal{Z}\left[J\right]=e^{iW\left[J\right]}
  = \int \mathcal{D}\mu \, e^{i S_0 + i S_J}
\end{equation}
Remember that $J_0$ and $J$ are sources for the same operators. The separation
into $S_0$ and $S_J$ is enforced by imposing the boundary conditions
eq.~(\ref{eq:L_0}) on $J_0$ as well as
\begin{equation}
  \label{eq:j_boundary_cond}
  \lim_{x \to \infty} J(x) = 0
\end{equation}

Functional integration in the susy environment demands control over the
complete set of susy variables \cite{susy}.  Since finding the path measure
$\mathcal{D}\mu$ does not directly interfere with what follows, we discuss its
derivation together with BRS gauge fixing in the appendix. We note that the
procedure adopted eliminates ab initio the auxiliary components of the vector
superfield $V$ and imposes after the elimination of all non propagating
components the persistent use of of the Wess - Zumino gauge. This is in
contrast to full susy chiral gauge fixing \cite{Gris}.

In the context of eq.~(\ref{eq:generating_functional}) the notion of susy
demands a twofold interpretation. By extrinsic susy we mean the full
supersymmetry the action $S_0 + S_J$ has by construction. Intrinsic susy is
the invariance of the highest component $L$ of the chiral superfield $\Phi$.
Choosing appropriate sources in $J$ will break intrinsic susy because the
algebra no longer closes into $L$. Extrinsic susy however remains unbroken for
arbitrary external fields $J$.

The generating functional for the connected Green functions $W\left[J\right]$
obeys the usual relations
\begin{equation}
  \label{eq:cl_fields}
  \frac{\delta W}{\delta j(x)} = \frac{1}{2} \langle L(x) \rangle =
  \frac{1}{2} L_{cl}(x), \qquad
  \frac{\delta W}{\delta m(x)} = -\frac{1}{2} \langle z(x) \rangle =
  -\frac{1}{2} z_{cl}(x) 
\end{equation}

The effective action $\Gamma\left[\Phi\right]$ is given by the Legendre
transform of $W\left[ J \right]$
\begin{equation}
  \label{eq:effective_action}
  \Gamma\left[\Phi\right]
  = \re \int d^4x \left(L_{\text{cl}}(x) j(x) - z_{\text{cl}}(x) m(x)\right) -
  W\left[J\right] 
\end{equation}
The sources that create prescribed classical fields $L_{\text{cl}}$ and
$z_{\text{cl}}$ are given by the functional derivatives of the effective
action with respect to the fields
\begin{equation}
  \label{eq:source_to_given_fields}
  \frac{\delta \Gamma}{\delta L_{\text{cl}}(x)} = \frac{1}{2} j(x), \qquad
  \frac{\delta \Gamma}{\delta z_{\text{cl}}(x)} = - \frac{1}{2} m(x)
\end{equation}
The relations in eqs.~(\ref{eq:cl_fields}) and
(\ref{eq:source_to_given_fields}) are equilibrium conditions for an
infinitesimal volume located at $x$. In other words, the system is in
equilibrium only if both sets of equations are valid simultaneously for all
$x$. To go to the static limit we choose the external fields $j,m$ to be
almost constant inside a small sub-volume $V_{\text{sub}}$ of the spacetime
$V$ and vanishing on the complement $V \setminus V_{\text{sub}}$ according to
eq.~(\ref{eq:j_boundary_cond}). Then we take the infinite volume limit
$V_{\text{sub}} \subset V \to \infty$. Technically this is equivalent to
choosing new boundary conditions
\begin{equation}
  \label{eq:j_to_j_inf}
  \lim_{x \to \infty} J(x) =
  - \vartheta^2 m_\infty + \frac{1}{g_\infty^2}-i \frac{\theta_\infty}{8\pi^2}
\end{equation}
Comparing with eqs.~(\ref{eq:L_0}) and (\ref{eq:j_boundary_cond}) we see, that
the boundary values $g_\infty^2,\theta_\infty$ and $m_\infty$ can be absorbed
into $S_0$ by the redefinition
\begin{equation}
  \label{eq:J_0_redef}
  J_0 \to -\vartheta^2 m_\infty
  + \frac{1}{2} \left(
    \frac{1}{g_0^2} + \frac{1}{g_\infty^2}
  - i \frac{(\theta_0 + \theta_\infty)}{8 \pi^2} \right)
\end{equation}
Transferring the boundary values from the source $J$ to the quantum mechanical
Lagrangian modifies the definition of the latter relative to
eq.~(\ref{eq:L_0}). Together with renormalization this amounts to the
substitutions
\begin{align}
  \label{eq:new_sources}
  J_0 &\to -\vartheta^2 m_{\text{R}}
  + \frac{1}{2} \left( \frac{1}{g_{\text{R}}^2}
  - i \frac{\theta_{\text{R}}}{8 \pi^2} \right) \notag \\ 
  J &\to 
  -\vartheta^2 (m(x)-m_{\text{R}}) + \frac{1}{2}
  \left( \frac{1}{g^2(x)} - \frac{1}{g_{\text{R}}^2} 
    -i \frac{(\theta(x) - \theta_{\text{R}})}{8 \pi^2}\right)
\end{align}
The subscript R refers to the renormalization group invariant couplings. Since
$g_{\text{R}}, \theta_{\text{R}}$ are constants we drop the surface term
proportional to $\partial_\mu(\Lambda^a\sigma^\mu\bar{\Lambda}^a)$ and find the
redefined Lagrangian
\begin{eqnarray}
  \label{eq:L_prime}
  \mathcal{L}_0' &=& 
  - \frac{1}{4g_{\text{R}}^2} F^a_{\mu\nu} F^{a\mu\nu}
  + \frac{\theta_{\text{R}}}{32\pi^2} F^a_{\mu\nu} \tilde{F}^{a\mu\nu}
  + \frac{1}{2} \Lambda^a i \overset{\leftrightarrow}{D}^{ab}
  \bar{\Lambda}^b \\
  &&- \frac{1}{2} m_{1\text{R}} (\Lambda^{a\alpha}\Lambda^a_\alpha 
  + \bar{\Lambda}^a_{\dot{\alpha}}\bar{\Lambda}^{a\dot{\alpha}})
  - \frac{i}{2} m_{2\text{R}} (\Lambda^{a\alpha}\Lambda^a_\alpha 
  - \bar{\Lambda}^a_{\dot{\alpha}}\bar{\Lambda}^{a\dot{\alpha}}) \nonumber
\end{eqnarray}
The above equation shows, that extrinsic susy incorporates in the quest of a
minimal set of equilibrium conditions an ensemble of quantum mechanical
Lagrangians with arbitrary complex external gaugino masses
$m_{1\text{R}},m_{2\text{R}}$. These mass terms break intrinsic susy. The
question to answer is, whether or not there remains a trace of this breaking
in the limit $m_{1\text{R}},m_{2\text{R}} \to 0$.

In order to construct an effective action simultaneously satisfying the
relations eqs.~(\ref{eq:cl_fields}, \ref{eq:effective_action},
\ref{eq:source_to_given_fields}) we extend the above procedure to arbitrary
combinations of sources $J$ inside $V_{\text{sub}}$ and $J'$ on the complement
$V \setminus V_{\text{sub}}$. Taking the infinite volume limit for all
possible combinations of sources and sub-volumes finally leads to the
effective action via the relations
\begin{equation}
  \label{eq:eff_action_from_exploration}
  \frac{\partial}{\partial \Phi} \Gamma(\Phi;J') = J \leftrightarrow J = J'
  \quad \text{and} \quad J = J' \to 0
\end{equation}
These operations are to be performed from left to right and in all permuted
sequences. Demanding $J = J'$ in the last step incorporates the equilibrium
condition for the two volumes in contact and involves the transfer of the
boundary conditions to the quantum mechanical Lagrangian according to
eqs.~(\ref{eq:J_0_redef} - \ref{eq:L_prime}).

Performing the infinite volume limit in all conceivable ways serves to explore
the effective action in the thermodynamic limit, which is in principle reached
herewith. Of course we cannot analytically go to this limit. Nevertheless it
turns out to be a useful gedanken-experiment, since extrinsic susy together
with the equilibrium conditions found are strong enough to imply relations
between different kinds of condensates in the above limit.

Note that in the thermodynamical limit the effective action plays the role of
the inner potential of the theory. Therefore we use the two notions
interchangeable in what follows.




\section{The chiral anomaly and the effective action}
\label{sec:chiral_anomaly}

Up to now the form of the effective action or inner potential in the
thermodynamical limit is only restricted by the validity of extrinsic susy. In
this section we will show, that in this limit the chiral invariance is
restored, leading to further restrictions on the K\"ahler and superpotential
that describe the inner potential. 

To show that the renormalized quantities defined in eq.~(\ref{eq:L_prime})
have a physical meaning beyond perturbation theory we first have to show the
existence of a renormalization scheme that leaves the chiral and trace
anomalies in the same multiplet to all orders in perturbation theory.

The structure of the axial current anomaly to one loop order is
\begin{equation}
  \label{eq:general_chiral_anomaly}
  \partial_\mu(\Lambda^a\sigma^\mu\bar{\Lambda}^a) = 2 C_2(G)
  \frac{1}{8\pi^2} \left( \frac{1}{4} F^a_{\mu\nu}
    \tilde{F}^{a\mu\nu} \right)
  = 2 C_2(G) \ch_2(G)
\end{equation}
In parallel the trace of the energy momentum tensor satisfies the anomalous
Ward identity to all orders \cite{trace}
\begin{equation}
  \label{eq:trace_anomaly}
  \begin{gathered}
    \vartheta^\mu_\mu = -3 C_2(G) \frac{1}{8\pi^2}
    \frac{\beta(g)}{\beta_{\text{1 loop}}(g)}
    \left( \frac{1}{4} F^a_{\mu\nu} F^{a\mu\nu} \right) \\
    \frac{\beta(g)}{\beta_{\text{1 loop}}(g)} =
    1 + \frac{1}{b_1}\sum_{n=2}^\infty b_n \kappa^n =
    1 + 2 C_2(G)\kappa + \cdots
  \end{gathered}
\end{equation}
where the rationalized coupling is defined to be $\kappa =
(\frac{g}{4\pi})^2$.  Provided $\beta_{\text{1 loop}} \neq 0$ the higher loop
corrections simply rescale the the renormalization group non-invariant
operator $\frac{1}{4g^2} F^a_{\mu\nu} F^{a\mu\nu}$ normalized to its off
shell, $\mu$ scaled two gauge boson matrix element being unity, such as to
render the product renormalization group invariant. At two and higher loop
level the renormalized operators $\Lambda^a\sigma^\mu\bar{\Lambda}^a$ and
$\ch_2(F)$ develop anomalous dimension functions $\gamma(\kappa) $
\cite{Kodei}, identical by the Adler Bardeen non-renormalization theorem. This
leads to a modification of the anomalous chiral Ward identity in
eq.~(\ref{eq:general_chiral_anomaly}) analogous to the higher loop
modifications of the trace anomaly. Keeping the chiral current renormalization
group invariant and defining the non-invariant operator $\frac{1}{4g^2}
F^a_{\mu\nu} \tilde{F}^{a\mu\nu}$ through the analogous normalization
procedure to its parity partner $\frac{1}{4g^2} F^a_{\mu\nu} F^{a\mu\nu}$,
\begin{equation}
  \label{eq:normalizing_j}
  \begin{gathered}
    \partial_\mu(\Lambda^a\sigma^\mu\bar{\Lambda}^a) = 2 C_2(G)
    \frac{1}{8\pi^2} \varepsilon(\kappa) \left( \frac{1}{4} F^a_{\mu\nu}
      \tilde{F}^{a\mu\nu} \right), \\
    \varepsilon(\kappa) = \exp \int_0^\kappa dk\frac{\gamma(k)}{b(k)},
    b(\kappa)= -\beta(g) g
  \end{gathered}
\end{equation}
it follows \cite{PMspin} that the two field strength bilinears, related by
susy, are identically renormalized to two loop order.  This fact has been
discussed in \cite{Novsvz} and using superspace integration techniques in
conjunction with dimensional reduction in \cite{Grisz}. Thus susy covariance
extended to the full set of superconformal, partially anomalous Ward
identities ensures the existence of a renormalization scheme with the all
order ultraviolet property
\begin{equation}
  \label{eq:all_order_identity}
  \frac{\beta(g)}{\beta_{\text{1 loop}}(g)} = \varepsilon(\kappa) 
  \quad \text{to all orders in $g$}
\end{equation}
Since in this framework $b_1 > 0$, the renormalization effects beyond one loop
simply lead to a redefinition of perturbatively renormalization group
invariant composite operators. For the right hand side of the chiral Ward
identity eq.~(\ref{eq:normalizing_j}) the all order framework is clearly
insufficient due to the exact quantization condition pertinent to the Chern
character $\frac{1}{8\pi^2} \varepsilon(\kappa) \left( \frac{1}{4}
  F^a_{\mu\nu} \tilde{F}^{a\mu\nu} \right)$ when continued to Euclidean space.

The existence of a renormalization scheme with the property
eq.~(\ref{eq:all_order_identity}) guarantees that the renormalized quantities
defined in eq.~(\ref{eq:L_prime}) have a physical meaning beyond perturbation
theory.

Using the renormalization group invariant quantities defined in
eq.~(\ref{eq:L_prime}) the chiral anomaly of the pure SYM theory in the
thermodynamic limit is given by
\begin{equation}
  \label{eq:chiral_anomaly}
  \partial_\mu(\Lambda^a\sigma^\mu\bar{\Lambda}^a) = 
  2 C_2(G) \frac{1}{8 \pi^2} \left( \frac{1}{4} F^a_{\mu\nu}
    \tilde{F}^{a\mu\nu} \right)
  - i (m_{\text{R}}\Lambda^{a\alpha}\Lambda^a_\alpha -
  \text{h.c})
\end{equation}
Thus the generating functional eq.~(\ref{eq:generating_functional}) satisfies
the relation
\begin{equation}
  \label{eq:equilibrium_for_Z}
  2 \left(
    \nu \frac{\partial}{\partial\theta} + \frac{\partial}{\partial \arg m}
  \right) \mathcal{Z} = 0 \quad \rightarrow \quad
  \mathcal{Z}=\mathcal{Z}\left(\frac{\theta}{\nu}-\arg m\right)
\end{equation}
where $\nu = C_2(G)$.  The discrete symmetry $Z_\nu$, 
\begin{equation}
  \label{eq:discrete_Z_nu}
  \theta \to \theta + 2\pi \frac{r}{\nu}
  \quad \text{for $r=0,1,2,\ldots,\nu-1$}
\end{equation}
of the generating functional $\mathcal{Z}$ is related to fixed time large
gauge transformations. Therefore we do not allow the $Z_\nu$ symmetry to be
broken spontaneously. This point has been discussed for $\nu \to \infty$ in
\cite{LeuSmi}. In the thermodynamic limit our analysis agrees.

For simplicity we drop the subscript R on the components of $J_0$ with the
understanding that $J_0$ always implies the use of the renormalization group
invariant quantities.

Using the composite variables $\chi, \bar{\chi}$ defined through the
relation
\begin{align}
  \label{eq:chi_chibar}
  \chi &= e^{-8\pi^2 j_0} = |\chi|e^{i\theta}, \qquad
  |\chi|=e^{-\frac{8\pi^2}{g^2}} \\
  \chi_{1/\nu} &= |\chi_{1/\nu}|e^{i\frac{\theta}{\nu}} \mod Z_\nu, \qquad
  |\chi_{1/\nu}|=e^{-\frac{8\pi^2}{\nu g^2}}
\end{align}
we see, that the phase dependence of $\Gamma\left[\Phi\right]$ is through the
products
\begin{equation}
  \label{eq:phase_dep}
  \bar{\chi}_{1/\nu}m, \qquad \chi_{1/\nu}\bar{m}, \qquad
  \chi_{1/\nu}z \quad \text{and} \quad \bar{\chi}_{1/\nu}\bar{z}
\end{equation}

The equilibrium conditions in eq.~(\ref{eq:equilibrium_for_Z}) with respect to
the imaginary part of the source $j,j'$ are tantamount to determine the
minimum of $\Gamma\left[\Phi\right]$ with respect to the boundary values
$\theta,\theta'$, as discussed in \cite{PMeta}. This minimum condition relaxes
$\theta,\theta'$ to the values
\begin{equation}
  \label{eq:theta_relaxed}
  \theta - \theta' = \nu \arg m \mod Z_\nu
\end{equation}
guaranteed by CPT invariance. To exclude nontrivial minima in the relative
phase $\Delta\theta = \theta - \theta' - \nu \arg m$ we remark that in the
Euclidean case
\begin{equation}
  \label{eq:abs_min}
  e^{-\Gamma_{\text{eucl.}}} \sim e^{+\cos{(\Delta\theta|z_{\text{cl}}|)}}
\end{equation}
Therefore the minima of $\Gamma$ are at $\Delta\theta=0 \mod 2\pi$ and are
absolute minima. Thus the $\theta,\theta'$ angles always relax in the
thermodynamic limit, allowing us to eliminate the variables dual to $j,j'$
first and to restrict the discussion to the remaining variables related to
$m$.

The relaxation of the $\theta$ angles restores the CP invariance when all
equilibrium conditions are met. Equivalently this can be interpreted as
restoration of the anomalous chiral symmetry in the thermodynamical limit.




\section{Linking spontaneous breaking of susy and restored chiral symmetry}
\label{sec:link_susybr_chiralbr}

In the last step we collect the knowledge of the form of the inner potential
restricted by extrinsic susy together with the restrictions from the restored
chiral invariance and the equilibrium conditions to derive relations between
the formation of the gaugino and gauge boson condensates.

As a consequence of the relaxation of the external variable $j,j'$ discussed
at the end of the previous section, the equilibrium conditions in
eqs.~(\ref{eq:cl_fields}) and (\ref{eq:source_to_given_fields}) become
\begin{equation}
  \label{eq:equil_for_Gamma}
  \frac{\partial}{\partial L_{\text{cl}}(x)}\Gamma(\Phi;J') = 
  \frac{1}{2} j(x) \rightarrow 0
\end{equation}
The above relation is the appropriate one for $L_{\text{cl}}$ and
$\bar{L}_{\text{cl}}$ being auxiliary fields of respective chiral multiplets
and the associated potential $\Gamma$ being minimized in the process of
eliminating them.

As discussed earlier extrinsic susy is strictly valid all the way and
restricts the effective action to be of the form
\begin{equation}
  \label{eq:form_of_eff_action}
  \Gamma\left( L_{\text{cl}}, z_{\text{cl}},
    \bar{L}_{\text{cl}}, \bar{z}_{\text{cl}}; m', \bar{m}' \right) = 
  - \bar{L}_{\text{cl}} K_{z\bar{z}} L_{\text{cl}}
  + \left (L_{\text{cl}} W_{z} +
    \text{h.c.} \right)
\end{equation}
where the factors $\chi_{1/\nu}$ and $\bar{\chi}_{1/\nu}$ now are sub-summed in
the variables $z_{\text{cl}}$ and $\bar{z}_{\text{cl}}$.  The functions
$K(z_{\text{cl}}, \bar{z}_{\text{cl}}; m', \bar{m}')$ and $W(z_{\text{cl}};
m', \bar{m}')$ denote the K\"ahler and super potential respectively. The
K\"ahler metric is given by $K_{z\bar{z}} = \frac{\partial^2 K}{\partial
  z_{\text{cl}} \partial \bar{z}_{\text{cl}}}$ whereas $W_z = \frac{\partial
  W}{\partial z_{\text{cl}}}$.

After transferring the boundary values of $m'$ to the quantum Lagrangian the
equilibrium value of $m'$ is zero and the equilibrium condition becomes
\begin{equation}
  \label{eq:final_equilibrium_cond}
  \frac{\partial\Gamma}{\partial L_{\text{cl}}} = 0 
  \quad \text{with} \quad
  \Gamma = \Gamma\left(L_{\text{cl}}, z_{\text{cl}},
    \bar{L}_{\text{cl}}, \bar{z}_{\text{cl}}; m, \bar{m}\right)
\end{equation}
where now the effective action is to be calculated with fixed Lagrangian
masses $m$ and for prescribed constant values of $z_{\text{cl}}$.
 
Eliminating the auxiliary fields $L_{\text{cl}}$ and $\bar{L}_{\text{cl}}$
according to eq.~(\ref{eq:final_equilibrium_cond}) we obtain 
\begin{equation}
  \label{eq:eff_action_L_eliminated}
  \bar{L}_{\text{cl}} = \frac{W_z}{K_{z\bar{z}}} \longrightarrow
  \Gamma = \frac{\left| W_z \right|^2 }{K_{z \bar{z}}}
\end{equation}

The superpotential is an analytic function of its natural variable
$z_{\text{cl}}$ and parametrically depends on both $m$ and $\bar{m}$.  The
induced K\"ahler metric must be positive to ensure stability of large volume
fluctuations. This is the positivity property of the internal potential that
is well known from semi classical approximations.

The relaxation of $\theta$ angles discussed at the end of section
\ref{sec:chiral_anomaly} implies the restored chiral invariance of the inner
potential under the phase rotations
\begin{equation}
  \label{eq:phase_rotations}
  z_{\text{cl}} \to e^{-i \xi} z_{\text{cl}}, \quad m \to e^{+i \xi} m \quad
  \text{ and c.c.}  
\end{equation}
assigning chiral weights 1 and -1 to $z_{\text{cl}}$ and $m$. Hence the
superpotential must have a well defined chiral charge whereas the K\"ahler
potential must be neutral.

Finally we relax the gaugino masses $m \to 0$,
\begin{equation}
  \label{eq:m_0_limit}
  \Gamma(|z_{\text{cl}}|^2; zm, \bar{zm}, |m|^2) \to
  \Gamma(|z_{\text{cl}}|^2;0,0,0) 
\end{equation}
It follows from eq.~(\ref{eq:m_0_limit}) that the potential $\Gamma$ attains
its minimum along a circle in the complex $z_{\text{cl}}$ plane. However the
only circle where $\bar{L}_{\text{cl}} \propto W_z$ can vanish is at the
origin $z_{\text{cl}}=0$. Therefore the only consistent solution with no
spontaneous breaking of susy does not admit any condensates. Then neither
gauginos nor gauge bosons can be confined, since a 'wall' to reverse their
chirality upon reflection is absent.

The remaining alternative of nontrivial condensate formation --- short of
spontaneous breaking of gauge invariance --- links the gaugino condensates
$z_{\text{cl}}, \bar{z}_{\text{cl}}$ with their gauge boson partners
$L_{\text{cl}}, \bar{L}_{\text{cl}}$. The latter spontaneously break susy at a
positive value of $\Gamma$. 

Let us discuss the form of the K\"ahler and super potential for this last
case. The inner potential is nonzero and positive at its minimum. Thus $W_z$
is nonzero, in which case the function $W=W(z_{\text{cl}})$ is invertible.
After the change of coordinates $z_{\text{cl}}=z_{\text{cl}}(W)$ the K\"ahler
potential eq.~(\ref{eq:eff_action_L_eliminated}) becomes
\begin{equation}
  \label{eq:eff_action_new}
  K_{W\bar{W}} = \frac{1}{\Gamma}
\end{equation}
where $K$ and $\Gamma$ depend on $W$ and $\bar{W}$. With the substitution $r =
\sqrt{W\bar{W}}$ eq.~(\ref{eq:eff_action_new}) is the radial part of the
inhomogeneous 2-dimensional Laplace equation
\begin{equation}
  \label{eq:Laplace}
  \left(\frac{d^2}{dr^2}+\frac{1}{r}\frac{d}{dr}\right) K(r) 
  = \frac{4}{\Gamma(r)}
\end{equation}
Notice that the source term $\frac{4}{\Gamma(r)}$ has a finite maximum at the
minimum of $\Gamma$ and goes to finite boundary values at zero and infinity.
Given $\Gamma$ we can solve eq.~(\ref{eq:Laplace}) for $K$,
\begin{equation}
  \label{eq:Kaehler_from_eff_action}
  K(r) =
  k + \int_0^{r} \frac{dr'}{r'} \int_0^{r'} dr'' \frac{4 r''}{\Gamma(r'')}
\end{equation}
The equations derived so far are all structural equations coming from
symmetries. The actual form of the effective potential can not be derived from
symmetries. It represents the genuine dynamics of the system.

Finally we emphasize that the conclusions reached are not without consequences
for more general supersymmetric theories.




\section{Summary and conclusions}
\label{sec:conclusion}

In the thermodynamical limit of a pure SYM theory the form of the effective
action is restricted by the validity of extrinsic susy. The restored chiral
invariance valid in this limit further restricts the K\"ahler metric and
superpotential describing the effective action. The equilibrium conditions for
the SYM system in contact with external sources driving the operators in the
Lagrangian as well as gaugino mass terms link the condensate of the gaugino to
the condensate of the Lagrangian and vice versa. The remaining alternatives
are: The gaugino condensate is zero and there are no gauge boson condensates,
or the gaugino condensate is linked to the gauge boson condensate. Assuming
confinement susy must be broken.

This result is in contradiction to the conclusion in \cite{witten_index} since
we do not meet the assumption of trivial boundary condition for the various
fields while deriving the thermodynamical limit.




\begin{appendix}
  
\section{Gauge fixing and extended functional measure}
\label{sec:gauge_fixing}

To derive the functional measure eq.~(\ref{eq:generating_functional}) we start
from the complexified superfield $V$, containing unconstrained components. The
extension of $V$ to unconstrained complex components allows us to extend the
Lorentz group, i.e.~its covering group $SL(2,C)$ to its complex form $SL(2,C)
\times SL(2,C)$ accompanied by the extension of configuration space
coordinates $x^\mu$ to complex values $z^\mu$. In addition we also extend the
fermionic coordinates $\vartheta^\alpha,\bar{\vartheta}_{\dot{\alpha}}$ of
$N=1$ superspace to $\vartheta^\alpha,\widetilde{\vartheta}_{\dot{\alpha}}$,
where $\widetilde{\vartheta}$ no longer is related to $\vartheta$ by complex
conjugation.

The now unrelated susy covariant derivatives become
\begin{equation}
  \label{eq:a_susy_cov_deriv}
  D_{\alpha} = \partial_{\alpha} - \frac{1}{2}
  i \partial_{\alpha \dot{\beta}} \widetilde{\vartheta}^{\dot{\beta}}, \qquad
  \widetilde{D}_{\dot{\beta}} = - \widetilde{\partial}_{\dot{\beta}} 
  + \frac{1}{2} \vartheta^{\alpha}
  i \partial_{\alpha \dot{\beta}}
\end{equation}
We can now perform gauge fixing \`{a} la BRS using throughout complexified
fields chiral with respect to the extended Grassmann variables
\begin{equation}
  \label{eq:a_extended_grassmann}
  \Theta^A = \left( \vartheta^\alpha, \widetilde{\vartheta}_{\dot{\alpha}}
  \right), \quad
  \partial_A = \left( \partial_\alpha, \widetilde{\partial}^{\dot{\alpha}}
  \right), \quad
  \left\{ \partial_A, \Theta^B \right\} = \delta_A^B
\end{equation}
To this end we use the complexified connections
\begin{equation}
  \label{eq:a_complex_connection}
  w_{\alpha} 
  = e^{-V} D_{\alpha} e^V, \qquad
  w'_{\dot{\beta}} = e^{-V} \widetilde{D}_{\dot{\beta}} e^V
\end{equation}
The connections in eq.~(\ref{eq:a_complex_connection}) only involve the
restricted pairs $\vartheta_{\alpha}$ and
$\widetilde{\vartheta}_{\dot{\beta}}$ separately and satisfy the relations
\begin{equation}
  \label{eq:a_complex_connection_relation}
  \begin{split}
    & D_\alpha w_\beta + D_\beta w_\alpha 
    + \left\{ w_\alpha, w_\beta \right\} = 0 \\
    & \widetilde{D}_{\dot{\alpha}} w'_{\dot{\beta}} 
    + \widetilde{D}_{\dot{\beta}} w'_{\dot{\alpha}} 
    + \left\{ w'_{\dot{\alpha}} ,
      w'_{\dot{\beta}} \right\} = 0
  \end{split}
\end{equation}
Next we set the specific constraints  
\begin{equation}
  \label{eq:a_constraints}
  N = D^\alpha w_\alpha, \quad
  N' = \widetilde{D}_{\dot{\beta}} w^{'\dot{\beta}}, \quad
  \mathcal{N} = N + N'
\end{equation}
The combined chiral field $\mathcal{N}$ is a hermitian field for restricted
hermitian values of $V$ and real space time variables $z^\mu$ in physical
space time.

The gauge is fixed by choosing a fermionic vector field $c$ with components in
the same basis as $V$ but with opposite fermion parity relative to those in
$V$. The field $c$ is to be interpreted in complexified extension.
\begin{equation}
  \label{eq:a_cov_deriv_on_c}
  \begin{split}
    \nabla_{\alpha} c &= D_\alpha c + 
    \left\{ w_\alpha , c \right\}  \\
    \widetilde{\nabla}_{\dot{\beta}} c &= \widetilde{D}_{\dot{\beta}} c + 
    \left\{  w'_{\dot{\beta}} , c \right\}
  \end{split}
\end{equation}

The fermionic BRS operator $\mathcal{S}$ is defined as operating on
$w_\alpha$, $w'_{\dot{\beta}}$ and $c$ like
\begin{equation}
  \label{eq:a_brs_operator_defined}
  {\mathcal{S}} 
  \begin{pmatrix}
    w_\alpha 
    \\
    w'_{\dot{\beta}}
  \end{pmatrix}
  =
  \begin{pmatrix}
    \nabla_\alpha
    \\
    \widetilde{\nabla}_{\dot{\beta}}
  \end{pmatrix}
  c, \quad
  \mathcal{S} c = c^2, \quad
  \mathcal{S}^2 = 0
\end{equation}
The nilpotency of $\mathcal{S}$ follows from its fermionic properties
\begin{equation}
  \label{eq:a_brs_fermionic_properties}
  \begin{split}
    & {\mathcal{S}} 
    D_{\alpha} 
    = -
    D_{\alpha} 
    {\mathcal{S}}, \quad 
    {\mathcal{S}} 
    \widetilde{D}_{\dot{\beta}}
    = -
    \widetilde{D}_{\dot{\beta}}
    {\mathcal{S}} 
    \\
    & {\mathcal{S}} 
    \left ( f_{1} f_{2} \right ) =
    \left ( 
      {\mathcal{S}} 
      f_{1} 
    \right ) 
    f_{2} 
    -
    f_{1} 
    \left ( 
      {\mathcal{S}} 
      f_{2} 
    \right ) 
  \end{split}
\end{equation}
with respect to a pair of fermionic superfields $f_{1} , f_{2}$. 

The BRS operation is completed extending the nilpotent action of $\mathcal{S}$
to the fermionic and bosonic vector superfields $\widetilde{c}$ and
$\widetilde{b}$ respectively by the relations
\begin{equation}
  \label{eq:a_tilde_c_tilde_b_defined}
  \mathcal{S} \widetilde{c} = \widetilde{b}, \qquad 
  \mathcal{S} \widetilde{b} = 0
\end{equation}

In our complexified environment the fields $c$, $\widetilde{c}$ and
$\widetilde{b}$ are independent before all BRS operations are performed.
Subsequently they are subjected to reality constraints as shown below.

We form the gauge fixing superfields $f$ (fermi) and $g = \mathcal{S} f$ from
the constraint superfields $\mathcal{N}$ in eq.~(\ref{eq:a_constraints}) by
means of the Lie frame independent traces
\begin{equation}
  \label{eq:a_f_g_defined}
  f = \tr \widetilde{c}
  \left( \eta \widetilde{b} + \mathcal{N} \right), \qquad
  g = \tr
  \mathcal{S}
  \left[
    \widetilde{c}
    \left( \eta \widetilde{b} + {\mathcal{N}} \right)
  \right]
  \equiv 
  \mathcal{S} f
\end{equation}
In eq.~(\ref{eq:a_f_g_defined}) $\eta$ denotes a real parameter, defining
Fermi gauges for gauge bosons together with the gauge invariant part of the
action.

Finally we form the complex densities $F$ and $G$ associated with the
superfields $f$ and $g$ respectively
\begin{equation}
  \label{eq:a_F_G_defined}
  \begin{pmatrix}
    F \\ G
  \end{pmatrix}
  =
  \int d^2 \vartheta \,
  d^2 \widetilde{\vartheta} \,
  \begin{pmatrix}
    F \\ G
  \end{pmatrix}, \qquad
  G \equiv
  {\mathcal{S}} F
\end{equation}

The following steps yield the gauge extended functional measure,
which we shall denote $\mathcal{D} \mu$ :

\begin{enumerate}
  
\item Setting reality constraints.

  We project back on physical space time and impose reality (hermiticity)
  constraints on the fields $c,\, \widetilde{c},\, \widetilde{b}$
  \begin{equation}
    \label{eq:a_project_to_physical_space}
    \begin{split}
      & z^{\mu} \rightarrow x^{\mu} = ( t , \vec{x} ), \quad
      \widetilde{b} \rightarrow b = b^{*}, \quad
      \left ( c , \widetilde{c} \right ) \rightarrow
      \left ( c , c^{*} \right )
      \\
      & V \rightarrow \frac{1}{i} V, \quad
      V = V^{*}, \quad
      \left ( G , F \right ) \rightarrow
      \re \left ( G , F \right ) 
    \end{split}
  \end{equation}

\item Eliminating non-propagating fields.

  We define the gauge fixing action
  \begin{equation}
    \label{eq:a_gauge_fixing_action}
    S_{g.f.} =
    \int d^{4} x \,
    \re G, \qquad
    S_{g.f.} =
    S_{g.f.} (\, b , c^{*} , c , V\, )
  \end{equation}
  All components of $b$ and some components of the superfields $c^{*},\, c,\,
  V$ are non-propagating. These components shall be eliminated choosing an
  extremum of $S_{g.f.}$. Hereby we choose the Wess Zumino gauge for $V$,
  which at this stage is a preselection of non-propagating components.

\item Constructing the extended functional measure.

  Having eliminated non-propagating fields, we define the extended functional
  measure in physical space time
  \begin{equation}
    \label{eq:a_extended_functional_measure}
    \mathcal{D} \mu =
    \prod_y 
    \left(
      \mathcal{D} V_y
    \right)
    \left(
      \mathcal{D} c^*_y
    \right)
    \left(
      \mathcal{D} c_y
    \right)
    \exp i 
    \left(
      S_{\text{g.f.}}
    \right)_{\mbox{\tiny extr.}}
  \end{equation}
  In eq.~(\ref{eq:a_extended_functional_measure}) the functional measure for
  the gauge field $V$ extends over all four hermitian components of the gauge
  boson potentials and over the chiral gaugino fields $\lambda^{a}$
  \begin{equation}
    \label{eq:a_components_of_functional_measure}
    \mathcal{D} V =
    \prod \mathcal{D} V_\mu^a \,
    \mathcal{D} \lambda^{* a}_{\dot{\beta}} \,
    \mathcal{D} \lambda^a_\alpha
  \end{equation}

\end{enumerate}
The extended functional measure $\mathcal{D} \mu$ defined in
eq.~(\ref{eq:a_extended_functional_measure}) through steps 1.~-~3. above is
independent of the gauge invariant part of the action. The latter is of course
quite unique. Together they form the basis of ultraviolet regularization and
renormalization of the theory, which is hereby separated from the infrared or
thermodynamic limit.

\end{appendix}




\vspace*{0.1cm} 

\vspace*{0.3cm} 

\noindent
\large{\bf Acknowledgments}
\vspace*{0.3cm} 
 
\noindent
We acknowledge interesting discussions with B. Ananthanarayan.  One of us
(P.M.) thanks the theory group of CERN and the members of the Institute for
Theoretical Studies in Bangalore, India, where part of this work developed,
for their warm hospitality.











\end{document}